# Privacy attitudes and concerns in the digital lives of older adults:

# Westin's privacy attitude typology revisited

Isioma Elueze, Anabel Quan-Haase

Western University

Abstract

There is a growing literature on teenage and young adult users' attitudes toward and concerns about online privacy, yet little is known about older adults and their unique experiences. As older adults join the digital world in growing numbers, we need to gain a better understanding of how they experience and navigate online privacy. This paper fills this research gap by examining 40 in-depth interviews with older adults (65+) living in East York, Toronto. We found Westin's typology to be a useful starting point for understanding privacy attitudes and concerns in this demographic. We expand Westin's typology and distinguish five categories: fundamentalist, intense pragmatist, relaxed pragmatist, marginally concerned, and cynical expert. We find that older adults are not a homogenous group composed of privacy fundamentalists; rather, there is considerable variability in terms of their privacy attitudes, with only 13% being fundamentalists. We also identify a group of cynical experts who believe that online privacy breaches are inevitable. A large majority of older adults are marginally concerned, as they see their online participation as limited and harmless. Older adults were also grouped as either intense or relaxed pragmatists. We find that some privacy concerns are shared by older adults across several categories, the most common being spam, unauthorized access to personal information, and information misuse. We discuss theoretical implications based on the findings for our understanding of privacy in the context of older adults' digital lives and discuss implications for offering training appropriate for enhancing privacy literacy in this age group.

# Privacy attitudes and concerns in the digital lives of older adults: Westin's privacy attitude typology revisited

Privacy is acknowledged as a basic human need and, as such, is an important policy and research topic (Bartsch & Dienlin, 2016; Kezer, Sevi, Cemalcilar, & Baruh, 2016; Walrave, Vanwesenbeeck, & Heirman, 2012). Much of the literature on privacy and digital media has examined the privacy concerns of young users (Acquisti & Gross, 2006), the types of information they disclose (Livingstone, Ólafsson, & Staksrud, 2011), and the kinds of strategies they employ to protect their privacy (Young & Quan-Haase, 2013). A key difference between older adults and younger users is that privacy concerns can be a real barrier to older adults participating online and expanding their media repertoire (Yuan, Hussain, Hales, & Cotten, 2016), whereas for younger users the evidence suggests that this is not the case (Jiang, et al., 2016). Younger adults have consistently been shown to rely on digital media despite having numerous privacy concerns (Acquisti & Gross, 2006), a phenomenon described as the privacy paradox or dilemma (Barnes, 2006; Kokolakis, 2017; Young & Quan-Haase, 2013). Media portrayals of privacy breaches, hacking, and cyberattacks create a sense in older adults that the use of digital media is risky. These privacy concerns in turn can make older adults reluctant to adopt digital media. For instance, Olphert, Damodaran, and May (2005) found that privacy concerns were a central barrier to older adults' uptake of the internet. They reported that a high proportion of older adults expressed privacy concerns, which in turn reduced their overall time spent online and hindered non-users' adoption of the internet. Similarly, Ferreira, Sayago, and Blat (2017) found that privacy was a major barrier for older adults in Brazil taking up digital media, suggesting that feeling in control of their privacy when online could potentially increase digital media uptake among older Brazilians. Adding to their reluctance to adopt is the fact that older adults often perceive their own digital literacy as low in comparison to younger generations (Schreurs, Quan-Haase, & Martin, 2017) and therefore feel unequipped to use these technologies safely or assess risks appropriately. Despite these many concerns, older adults are adopting not only traditional digital media such as email, but also more interactive types including Facebook and Skype, even if to a lesser extent (Anderson & Perrin, 2017). While some older adults may be willing to interact on social media, their understanding of the affordances of these sites is limited (Quan-Haase, Williams, Kicevski, Elueze, & Wellman, 2018). The conundrum, of course, is that since older adults do not have as much expertise posting and interacting on social media and adjusting their privacy settings, they may be at a greater risk.

Considering the attention Westin's typology of privacy attitudes has received in prior literature, we employ his theoretical framework to inform the present study in the context of older adults. Using Westin's typology also facilitates making comparisons across studies that are based on different samples and different age ranges. Through an examination of 40 interviews with older adults (65+), we test and expand Westin's much-debated typology. We propose a revised typology that distinguishes five categories: fundamentalist, intense pragmatist, relaxed pragmatist, marginally concerned, and cynical expert. In addition, we investigate what kinds of privacy concerns older adults in each category have. The significance is to provide public policy insights into this social group, demonstrating the value of support through privacy literacy training and coupling adoption of digital media with risk mitigation strategies.

**Literature Review**

Much recent scholarly research has focused on privacy and how notions of privacy continue to evolve as a result of widespread adoption of digital media (Tsai et al., 2016; Tufekci, 2007). Dhir, Torsheim, Pallesen, and Andreassen (2017) suggest that young adults understand online privacy better because of their heavy reliance on digital media. Courtney (2008) found that the meaning of privacy varied widely among older adults (65 years and older) and included a desire to be alone, to control the information shared with others, to control access to one's personal property, and to protect oneself from identity theft. As a greater proportion of older adults go online (Anderson & Perrin, 2017), understanding this group's attitudes and concerns can help reduce anxiety and provide better privacy literacy.

**Theoretical framework: Westin's typology of privacy attitudes and older adults**

We build on Westin's theoretical framework to better understand the privacy attitudes and concerns of older adults. In addition to analyzing more than 120 privacy surveys held in the Privacy & American Business survey library, Alan Westin supervised about 45 national privacy surveys between 1979 and 2001 in the US (Bracy, 2013). Using results from these surveys, Westin categorized respondents into fundamentalist, pragmatist, and unconcerned (Kumaraguru & Cranor, 2005). Data were gathered from randomly selected members of the US population, and the classification is based on responses to a Likert-type scale. Although there were variations in the focus of the numerous surveys (e.g., consumer privacy, health information privacy, and e-commerce) as well as in the proportion of participants belonging to each privacy category, Westin's description of the tripartite categories remained stable.

Westin described each category in terms of differing privacy attitudes (see Table 1). Privacy fundamentalists are suspicious about anything that they perceive as a threat to their privacy and are unwilling to disclose their personal information. Pragmatists weigh the risks of giving out personal information against the potential rewards. Finally, unconcerned individuals are comfortable with sharing their information with organizations, believing that the information is generally safe. Westin's *"privacy on and off the Internet"* survey (Kumaraguru & Cranor, 2005; Westin, 2000) found that among adult Americans, 25% could be categorized as fundamentalists, 55% as pragmatists, and 20% as unconcerned. In sum, Westin's (2000) work provided a typology of online users and demonstrated individual differences in privacy attitudes. In the present study, Westin's typology informed the coding process and provided a baseline with which to compare the study findings.

<Table 1 here>

Despite its widespread application in privacy research (King, 2014; Motiwalla & Li, 2016), Westin's categorization has received criticism (e.g., Urban & Hoofnagle, 2014). Urban and Hoofnagle (2014) found that, in practice, many consumers made pragmatic decisions even when they were categorized as unconcerned. Motiwalla and Li (2016) found mixed support for Westin's three categories. They found comparable groupings in their sample, but variation in the distribution across categories: 25% were fundamentalists, 69% were pragmatists, and only 6% were unconcerned. This suggests that the percentage of individuals falling into each category may vary both over time and by type of population. In addition, Sheehan (2002) found that internet users who were 45 years of age or older made up the majority of the unconcerned or very concerned category, while younger users were more likely to be pragmatists. Supporting the idea that different populations could fall into different groupings, King (2014) found fundamentalists to be older, while privacy unconcerned were generally younger. Martin, Gupta, Wingreen, and Mills (2015) found mixed support for Westin's groupings. They found two types of pragmatists instead of one; the first type preferred to be notified of the use of their personal information prior to its being collected and also expected organizations to obtain consent for the use of their information. The second type of pragmatist viewed any collection of personal information as requiring prior consent, but once that information was collected this group was agreeable to having it disclosed to another person or organization if necessary. Martin et al.'s (2015) study found no evidence for an unconcerned category. Dupree, DeVries, Berry, and Lank (2016) found support for Westin's fundamentalist and marginally concerned groups, yet subdivided the pragmatists into three groups—lazy experts, technicians, and amateurs. The reviewed studies suggest that Westin's categories can serve as a starting point for privacy scholars and help in grouping users into different broad categories. They also suggest that age needs to be considered more closely, as older adults may not fall into a single category, but rather may constitute different categories.

RQ1: What privacy categories do older adults fall into?

**Online privacy concerns and older adults**

Users of digital media express a wide range of concerns regarding their privacy (Tsai et al., 2016); even users who evaluate their digital skill level as "high" report concerns, despite being better equipped to prevent privacy threats than users who report having "low" skills (Spake, Zachary Finney, & Joseph, 2011). Often reported privacy concerns include the misuse of personal information posted online (Barnes, 2006); lack of trust for banking online (Alhabash et al., 2015); and concerns about identity theft, fraud, or bullying online (Bartsch & Dienlin, 2016).

Much of the literature investigating online privacy concerns has looked at adolescents (Dhir, Kaur, Lonka, & Nieminen, 2016; Livingstone et al., 2011) and young adults (Van den Broeck et al., 2015; Walrave et al., 2012), drawing on data collected from conveniently available university students (e.g., Acquisti & Gross, 2006; Suh & Hargittai, 2015). Fewer studies have focused on older adults aged 65+ and their online privacy concerns (exceptions include for example Courtney, 2008; Kezer et al., 2016). Some studies have suggested age differences in online privacy concerns. For instance, an European Union study of children aged 9–16 found that 28% set their social network site profile to "public" and 13% included their address, suggesting their concerns about privacy are low (Livingstone, et al., 2011). In addition, Dhir et al. (2016) found that adolescents with more experience on Facebook have fewer online privacy concerns. This kind of laissez-faire attitude toward privacy has often led to the assumption that adolescents and young adults in comparison to older adults show few to no privacy concerns. To understand generational differences, Miltgen and Peyrat-Guillard (2014) conducted focus groups with young people (15–24) and adults (25–70) in seven European countries. The authors found that middle-aged respondents (45–60) perceive more privacy risks online and have a greater fear of privacy invasion compared to younger people (25–44).

RQ2: What kinds of privacy concerns do older adults in each of the categories have?

# Methods

## East York

We draw on interviews from the fourth wave of data collection that has taken place in East York since 1968 (Wellman, 1979; Wellman & Wortley, 1990; Wellman, et al., 2006). East York, previously an autonomous borough of metropolitan Toronto, is now part of the larger City of Toronto, the fourth-largest metropolitan area in North America. The population of East York has remained stable since the 1960s with about 100,000 inhabitants (Quan-Haase, et al., 2017). East York is, like much of Toronto, culturally diverse. In 2000, 53.5% of East York residents were Canadian-born and 45.1% were foreign-born (Statistics Canada, 2002). In 2000, the median income of East York inhabitants aged 15+ was $24,408, comparable to the whole of Ontario at $24,816 (Statistics Canada, 2002). The median age of East York residents was 37.4 in 2000, and 13.6% of individuals were over age 65. Like the rest of Toronto, East York has experienced the spread of apartment buildings that supplement its formerly predominantly single-family houses.

**Sample**

Our overall sample comprises 101 participants, 57 women and 44 men. The mean age of the sample is 59.5, ranging from 27 to 93 years of age. The current paper focuses on the subset of 40 older adult participants between the ages of 65 and 91 who completed the privacy questions (one older adult, aged 65+, was excluded because responses to the privacy questions were missing). The mean sample age was 73.4 (*S.D.*= 6.6). The sample consisted of 21 women and 19 men. Most of them were retired from their careers, and actively engaged with friends, relatives, community, and other contacts. Many respondents were culturally British-Canadian, and the diversity of the other participants precluded further analysis of cultural variation. Of our sample, 92% owned a computer, 87% had a mobile phone, and 18% owned a tablet. Yet of those who owned a mobile phone, most used it solely for emergencies and 22% reported texting. Email was employed by 83% of respondents, 35% used Facebook, and 33% used Skype. All but three of the older adults noted at least one form of internet use or online activity.

**Procedures**

To obtain a representative sample of East York residents, we based our sampling frame on a list of 2,321 randomly selected residents provided by Research House, a Toronto-based research firm. We randomly contacted 304 people on the list, of which 101 individuals agreed to participate, a response rate of 34%. Each selected individual received an information letter, approved by the University of Toronto's Research Ethics Board (see Quan-Haase et al., 2017, for procedural details).

**Data Analysis**

Data analysis was guided by the tenants of qualitative work (Charmaz, 2014) and specifically thematic analysis (Braun & Clarke, 2006) (Figure 1). The first author familiarized herself with the data and then generated an initial set of codes and memos based on five randomly selected transcripts (Charmaz, 2014). To ensure rigour and increase intracoder reliability, the first author read and coded another set of three randomly selected interviews while employing the same process of open coding that remained close to the data, including memo writing and reflection (McKechnie, Chabot, Dalmer, Julien, & Mabbott, 2016). Corbin and Strauss (2008) also stress the value of employing past frameworks for informing the data analysis process in qualitative work. The first author relied on Westin's tripartite grouping as a sensitizing tool. Responses to the privacy questions were read closely for each interview and based on respondents' key statements they were grouped into one of Westin's categories. Westin's descriptions for each category as summarized in Table 1 guided the coding and factors that were considered in assigning the older adults included statements regarding how private they considered themselves, their level of awareness of risks of online activities/information disclosure, and how reluctant they were about sharing information online. Any ambivalences were noted and additional coding took place. The coding process was also open to new insights, hence the emergence of new groupings. We found that the pragmatists group had some older adults who demonstrated an understanding of the trade-off between engaging in online activities and information disclosure, while there were others who appeared more relaxed in their privacy concerns but were not unconcerned, which led to splitting pragmatists into two groups.

After the groups were established, the privacy concerns of the older adults within each category were also coded. To strengthen credibility of the data analysis process, the first and second authors worked closely together to refine codes, review codes, and identify anomalies. Data analysis also included the exploration of negative cases or anomalies to further enrich the data (Genius, 2015). Following Houghton, Casey, Shaw, and Murphy (2013), we used thick description and supporting quotes throughout the findings to increase the trustworthiness of the data, and pseudonyms to protect the confidentiality of our interviewees.

<insert Figure 1 about here>

## Findings

### A typology of older adults' privacy attitudes

RQ1 investigated what categories older adults could be clustered into based on their attitudes toward online privacy. We found some support for Westin's tripartite typology, as we could categorize the older adults into fundamentalists, pragmatists, and marginally concerned. Pragmatists, however, were not a homogenous group; rather, two sub-groups emerged. Relaxed pragmatists considered themselves private in some ways and demonstrated some understanding of risks involved in information disclosure online, but were less reluctant than intense pragmatists to share their personal information. Furthermore, through our analysis of negative cases and anomalies, we identified two participants (P3 and P60) who did not fit well within Westin's framework. Its significant difference from the other groups led us to propose a fifth category: cynical expert. Because only two participants belonged to this group, it was small, yet coherent. Guided by Westin's framework, Figure 2 shows the criteria used for coding our participants into the five categories.

<Figure 2 around here>

We found that 57% of our East York older adults were pragmatists—of which 42% of the total sample were relaxed pragmatists and 15% were intense pragmatists—followed closely by 25% who were marginally concerned, 13% who were fundamentalists, and 5% who were cynical experts. Table 3 compares our findings to other research, demonstrating a distinction in how participants from our sample were distributed versus the participants' placement in previous studies.

<add Table 2 here>

**1. Fundamentalist (13%)**

Fundamentalists consider themselves to be very private. A guiding principle for them is that one's personal matters and information should be kept to oneself. Consequently, fundamentalists do not give out personal information, and some demonstrate additional concern regarding what may happen if their information gets collected. What particularly unnerved the fundamentalists was the ease with which organizations could access personal information online, and the prospect of falling prey to cybercrimes. Fundamentalists were generally very unwilling to share information about themselves online and also wondered what motivated others to do so, as exemplified by Paul:

*I would never reveal the sorts of things online—I can't imagine I ever would—that other people do, including posting pictures of themselves with their current partners or themselves nude, and so on. (P84, M, 81)*

Often fundamentalists attributed their unwillingness to disclose information online to their age:

*I don't post my birthday online. I don't put family information online. I don't post photos of grandchildren or anything… I think older people are more concerned than younger people. (P55, M, 68)*

And Sidney exemplifies how fundamentalists see their age as a vulnerability that others are looking to exploit:

*When you get to be our age, you hear more and more scams that are perpetrated onto seniors. You're lucky enough to be able to trigger or catch onto what's real and what's not so… I don't like to give too much out. (P26, W, 68)*

Although, none of them reported having had any negative experiences when using digital media, fundamentalists see the use of social media sites, online banking, and e-commerce as risky activities, where there is a high likelihood of privacy being breached.

*Basically people can have their identities taken. Your credit card can be compromised, your social security can be compromised. It doesn't take much for somebody to impersonate somebody. (Tom, P55, M, 68)*

Fundamentalists also thought social media users shared too much personal information, which they found both dangerous and distasteful. Because of their wariness, fundamentalists preferred to stay away from social media sites and chose not to engage in e-commerce or online banking.

*Too much. I mean I've had it blow up in so many people's faces that I can't even tell you and I think it's due to their own stupidity. I don't think it's probably a Facebook problem. Although Facebook has been hacked but I think due to their stupidity. (Rachel, P98, W, 72)*

**2. Intense Pragmatist (15%)**

Intense pragmatists were like fundamentalists in that they considered themselves to be private, and were aware of the risks involved in sharing too much personal information online. Even though they did not like giving out personal information online, they recognized the need to give out some information as part of a necessary trade-off between protection of privacy and need to engage in an activity in certain situations, as in the example of e-commerce. Acknowledging privacy risks when using digital media and an unwillingness to share too much information online, Anastasia explains her approach:

*Usually I say I don't want to be here [online] anymore unless it's something I'm really interested in. I don't do much buying over the internet but I have on occasion bought a few things there that I couldn't find. (P37, W, 66)*

Likewise, Mark expresses how he limits what information he shares, but makes these decisions based on the potential benefits he might obtain.

*I try to limit it. If I feel information might be useful to certain organizations I might allow it. It depends. (P80, M, 65)*

What further unified these participants were the strategies they employed to maintain their privacy online due to their awareness of the risks in disclosing information when using digital media. These strategies gave them a sense of efficacy in keeping control over their personal information. Unlike the fundamentalists, the intense pragmatists did not view their age as a factor in susceptibility to online scams. However, they expressed the opinion that older generations were overall more reserved in what they shared online than younger generations.

*Some of the younger ones like my friend, [Katherine], look upon me as a dinosaur as far as paying my bills [online] are concerned. They say, ‘Oh the convenience, it's so fabulous’, but I don't mind stepping to the bank and chatting up the tellers and paying my bills in person. (Veronika, P62, W, 73)*

**3. Relaxed Pragmatist (42%)**

Relaxed pragmatists were more varied as a group in the extent to which they valued privacy. The relaxed pragmatists considered themselves private in some ways, but less private in others. Like intense pragmatists, relaxed pragmatists felt that online privacy was worth maintaining and were not very willing to share information when engaging with digital media. They tended to weigh the benefits:

> *Because of these people in California, I kind of like keeping in touch with them. And you can…you do get a lot of junk but you don't have to look at it, you know? (Beverly, P2, W, 75)*

However, their familiarity with the associated risks was much more limited than that of the intense pragmatists:

> *Well, my credit card gives me some points and sometime that's good. It has nothing to do with your privacy. (P15, An Dung, M, 71)*

Their strategies were also less concrete, and were viewed as common sense rather than specific strategies.

> *No, I don't do anything to protect my privacy. I mean, everybody has a code, but you know, that's normal practice. (Joe, P7, M, 73)*

Some relaxed pragmatists did not share the same views as the fundamentalists or the intense pragmatists regarding the effect their age had on online safety. Maria expresses her belief that age is not a factor in terms of the risks, and states that she sees anyone as potentially being at risk when using digital media:

*I think anybody who could be, maybe a young woman, maybe a woman who didn't show a lot a confidence, maybe an older woman, maybe a … they might try it on anyone. I don't know but there's an awful lot of it out there. (Maria, P90, W, 82)*

**4. Marginally Concerned (25%)**

The marginally concerned participants were distinct in that they either did not consider themselves private, or they did, but believed that their online presence had little to do with their sense of privacy. Some noted that they did not feel concerned about divulging information about themselves to organizations, but were simply curious as to how it was used. This indifference was often linked to low awareness of the potential risks involved when using digital media. Marginally concerned participants believed they were not putting themselves at risk because their online engagement was of little relevance and generally consisted of harmless activities, as Benjamin stated:

*So most of these things are things that I've gone to, asking on Google … asking about arthritis or a medical condition. (P31, M, 80)*

Marginally concerned participants had an indifferent attitude toward maintaining online privacy, the potential risks associated with digital media use, and organizations collecting their personal information. Some marginally concerned, like Magdalena, thought they could prevent falling prey to privacy breaches by being cautious and clever:

*Couldn't care less. Nothing to hide. ... Can't fool me. I'm too smart (P73, Magdalena, W, 75)*

**5. Cynical Expert (5%)**

The cynical experts, Aaron and Devon, were both technologically adept and well-versed in various privacy risks associated with being online. While Devon considered himself private, Aaron was indifferent. Aaron was one of the three older adults in our sample who had a Twitter account, and Devon was on Facebook. Despite revealing more expertise regarding online privacy than the other participants in the sample, these cynical experts demonstrated little inclination to employ privacy strategies. They were cynical about how much a user could prevent organizations from collecting personal information, as Aaron states:

*In terms of organizations collecting information about me, if they're interested, I think it's inevitable. (Aaron, P3, M, 69)*

For cynical experts, their perception of technology and associated risks created a feeling of impotence about how much they could protect their data, particularly against organizations. For them, it was inevitable that organizations would collect and use data in ways that were impossible for them to foresee. They did not see their own age as a factor that either increased or decreased privacy risks; rather, they saw all users as being at the mercy of organizations. Simply stated by Devon,

*I mean, if they're going to monitor your emails, they're going to monitor your emails. (Devon, P60, M, 70)*

**Privacy concerns of older adults**

In RQ2, we investigated the privacy concerns of older adults in each of the five categories. We found that most online privacy concerns such as surveillance, scams, unauthorized access to personal information, identity theft, and information misuse were shared by participants in different categories. Concerns about intellectual property and knowledge theft were mentioned by only one participant and seem to be rather rare among this demographic. Within the groups, we found that marginally concerned individuals have very few concerns, while cynical experts have numerous concerns. Having more concerns than relaxed pragmatists, intense pragmatists not only worry about privacy, but also more actively protect it. Fundamentalists have many concerns, but protect themselves by not engaging online, to reduce anxiety.

**Fundamentalists**

Fundamentalists demonstrated heightened anxiety around online privacy and had many online privacy concerns, ranging from a fear of online scams to being hacked. In addition to the fear of being hacked, a concern shared by Olga and Paul, Sidney was concerned about being scammed; she inquired multiple times during the interview if there was a new scam out there that she needed to be aware of:

*Is there another scam out there I don't know about? (P26, W, 68)*

Paul demonstrated that fundamentalists typically do not want any of their private information to be known to other individuals or to organizations:

*I just don't want people to know even harmless things about me. (P84, M, 81)*

And Tom elaborated on why:

*I am wary about losing control [of] information about me and how it might be used. (P55, M, 68)*

**Intense Pragmatists**

Intense pragmatists had a range of concerns such as unauthorized access to their personal information. But intense pragmatists tried to stay a step ahead of concerns they had. For instance, for fear of unauthorized access to her information, Anastasia engaged minimally online:

*I guess it would be somewhere along the line… someone gets access to it that doesn't deserve to, so I try to sort of keep my business dealings out of the computer. (P37, W, 66)*

Also, the fear of potential illegal activities made Sterling do his online banking through an encrypted desktop computer:

*I do tons of banking through the internet, but that's out of an encrypted desktop computer. (P93, M, 70).*

**Relaxed Pragmatists**

Relaxed pragmatists expressed fewer concerns. Like intense pragmatists, relaxed pragmatists were also concerned about others having unauthorized information about them. For example, Maria believed that the information collected about her online could be misused:

*People getting your information to use for their own purposes. There's much more chicanery now than there ever was (P90, W, 82).*

According to relaxed pragmatists, websites wanted too much personal information that often was not directly related to the task or activity at hand. These concerns, however, did not preclude some relaxed pragmatists from engaging in e-commerce; rather, John observed the invasion of privacy as puzzling:

*Ordering a book or music online and they want to know details of how many children you have, where they live—that sort of thing. I'm somewhat mystified by that. (P101, M, 70).*

But unlike John, the fear of threats to their privacy could lead some relaxed pragmatists to stay away from using digital media. This tactic is illustrated by Michael, who was concerned about unauthorized access to his information:

*People accessing your accounts or personal information. I try to keep the amount of information to a minimum. I don't chat on the internet, let's put it that way. (P29, M, 91)*

**Marginally Concerned**

Those in the marginally concerned category expressed very few, if any, concerns. Six of the ten older adults in this category did not mention any kind of online privacy concern, two implied that they were careful with their information without mentioning any concerns, and two revealed privacy concerns related to (1) being scammed and (2) spam emails. Samar and Benjamin, two marginally concerned older adults, noted the following:

*Today my fear is lot of scams going on. (P83, W, 82)*

*I just don't want another bunch of mail coming in… I am going to get bombarded with junk mail for a long time. (P31, M, 80)*

**Cynical Experts**

Probably due to their skepticism concerning online privacy, this group had a high concern level. In addition to being concerned about junk email, Devon was concerned about surveillance by government organizations. He referred to disclosures by Edward Snowden (which were then in the news), and to an arrest linked to "an innocent" Facebook post:

*All this NSA [National Security Agency] stuff in the States, you know, and the guy that's in Russia right now? Snowden? And, the information he's releasing about all the information the US is gathering on people. Logging all their emails, their phone calls, their just about everything? There's no privacy down there at all anymore. (P60, M, 70)*

Similarly, Aaron (P3, M, 69) was concerned about unauthorized access by scammers to his information when engaged in online banking. Cynical experts were concerned, yet felt that not much could be done to prevent surveillance.

## Discussion

We investigated to what extent East York older adults would fall into Westin's typology and the kinds of privacy concerns raised in each category. Past studies have shown mixed support for Westin's categories (Woodruff et al., 2014), with some rejecting the typology and others extending it by adding new categories. The motivation for the study came from the finding that older adults are overly concerned about online privacy and these concerns result in them limiting their digital media use (Jiang et al., 2016). This would suggest that many older adults would belong to Westin's fundamentalist category. Our findings contradict these generalizations and show that older adults are not a homogenous group comprised of privacy fundamentalists; rather, there is considerable variability in terms of their privacy attitudes, and in fact only 13% were grouped as fundamentalists. For the older adult fundamentalists privacy was a high priority, and personal matters and information needed to remain completely private.

Our findings generally support Westin's tripartite typology. However, corroborating with past studies, we found that additional categories were needed. The revised typology includes five categories: fundamentalist, intense pragmatist, relaxed pragmatist, marginally concerned, and cynical expert. Like Martin et al. (2015), who further fragmented pragmatists into two groups, we also subdivided pragmatists into two groups because not all of them shared the same privacy attitudes. In addition to being less willing to share personal information online and a greater understanding of the risks associated with disclosing information, intense pragmatists had a repertoire of tactics that they utilized, making them more active in their privacy maintenance than the relaxed pragmatists. Relaxed pragmatists employed only basic, common-sense strategies to safeguard their privacy. The two pragmatist groups we identified are, however, distinct from Martin et al.'s two pragmatist groups. Both of Martin et al.'s pragmatist groups were willing to share information if they could give consent prior to their personal information being collected and used. By contrast, our pragmatists were less concerned regarding consent and differed in terms of their willingness to engage in privacy protection strategies.

We found support for Westin's marginally concerned group, which comprised older adults who did not consider themselves very private and did not mind sharing information with organizations and when using digital media. Often older adults in this group were limited internet users, had little understanding of online risks, or showed low to no concern about data misuse. Finally, we found that a small anomalous group emerged during coding, which we termed cynical experts. They were characterized by intense cynicism: according to them, nothing really could be done to protect their privacy against large corporations. This attitude corresponds with previous findings from Hoffman et al. (2016), who found that some users develop a sense of "privacy cynicism" as a response to feeling overwhelmed by privacy threats. Similarly, Hargittai and Marwick (2016) discuss "privacy apathy" in younger generations because "the necessity of using social media made some participants express resignation about privacy violations and a lack of ability to change this situation" (p. 3751). Older adults in the intense cynicism category felt that, given the various risks involved in using digital media, there was little to be done in terms of protecting their privacy and saw privacy loss as inevitable when adopting digital media.

Our findings show some overlap with previous distributions of individuals into categories, but also some differences (see Table 2). The size of the group of fundamentalists is somewhat comparable to other study findings, it is certainly not much higher than other study findings. In our study, because of their wariness, fundamentalists preferred to stay away from social media sites and chose not to engage in e-commerce or online banking. This shows that low privacy literacy combined with a desire for privacy may deter fundamentalists from adopting new types of digital media, even if they could benefit from them. We observed a greater proportion of marginally concerned individuals in our sample compared to four other studies (see Table 2), a study by Ackerman et al. (1999) reported about the same proportion. The high percentage was unexpected because, in addition to the portrayal of older adults as having high privacy concerns (Ferreira et al., 2017; Olphert et al., 2005), there is a general impression that older adults are wary of online privacy breaches. Future research can examine how stable the groupings are over time and to what extent the proportions are typical for older adults in comparison to other age brackets.

**Limitations and Future Research**

As digital technology continues to evolve quickly, and as new online threats emerge at a similar pace, our findings may need to be revisited. A key limitation of our study is the cross-sectional design, which does not allow us to examine how privacy attitudes and concerns may change over time. For example, as older adults spend more time online, they may learn to better protect their privacy. More fundamentally, to what extent are the findings a reflection of the participants' age or their membership in a specific generational cohort? Future studies can better disentangle these age effects. Finally, we identified privacy literacy as an important dimension of older adults' online experience. A relatively new stream of research is emerging that investigates online privacy literacy and how it can prevent privacy breaches and protect personal information (Bartsch & Dienlin, 2016; Li, 2018; Trepte et al., 2015). Future research can specifically examine the privacy literacy levels of older adults and how they can be increased through additional forms of engagement online as well as through formal (e.g., training programs) and informal learning (e.g., peers, family, and the media) (Schreurs, et al., 2017).

## Conclusion

Older adults were often reluctant to share information online because they saw their age group—seniors—as being a frequent target of scams and other cybercrimes (also see Park, 2013). The fact that email was used by many of the older adults may explain the high fear of receiving unsolicited emails including scams. Many mentioned that they were particularly vulnerable because of their low privacy literacy, which precluded them from taking the right steps to protect themselves from threats. Park (2013) found that the fact that "older users are less skillful than are younger people in privacy control creates a grim scenario in which they may be the worst victims of identity theft or related online crimes" (p. 231). This suggests that privacy literacy is an important aspect of the digital inclusion when looking at questions of digital inclusion for this age group. Unlike younger users, who continue to engage online despite privacy concerns, older adults' privacy concerns are a *real* barrier to both adopting and fully engaging with digital media. Additionally, if users feel that their online privacy is at risk regardless of their practices, they will be disinclined to utilize privacy protection strategies, placing them at a greater risk. Therefore, policy should not only focus on the educational aspect of online privacy measures, but also instill a sense of self-efficacy in older adults who feel as though such efforts may not be worthwhile.

## Acknowledgements

We are grateful to Carly Williams, Barry Wellman, Helen (Hua) Wang, Alice (Renwen) Zhang, Christian Beerman, and Rhonda McEwen for their collaboration, the editors and reviewers for their comments, the Social Sciences and Humanities Research Council of Canada (SSHRC) for financial support, and most importantly, the East Yorkers who invited us into their homes.

**Disclosure statement**

No potential conflict of interest was reported by the authors.

## Tables

Table 1. Westin's Privacy Attitude Typology

| Privacy Fundamentalists (25%) | Privacy Pragmatists (55%) | Privacy Unconcerned (20%) |
|---|---|---|
| -Highly value privacy.<br>-Believe that they own their personal information, and advocate for individuals to safeguard personal information from organizations, businesses, and governments.<br>-Support strong laws put in place by the government to secure privacy rights and regulate organizations when collecting personal information. | -Assess the benefits of disclosing personal information to organizations, businesses, and governments.<br>-Advocate for fair information practices, and give out their information to organizations that they trust.<br>-Individuals should be responsible for making decisions regarding information disclosure and support government. | -Do not understand the "privacy fuss."<br>-Not particularly bothered about privacy abuse.<br>-Willing to give out their information to organizations, governments, and businesses.<br>-Do not advocate for government legislation to protect individuals' privacy. |

Note: based on Westin (2000).

Table 2. Comparison of Findings from Older Adults to Other Studies based on Westin's (2000) Typology

| | Present paper | Dupree et al. (2016) | Consolvo et al. (2005) | Sheehan (2002) | Westin (2000) | Ackerman et al. (1999) |
|---|---|---|---|---|---|---|
| Marginally Concerned (Unconcerned) | 25% | 16% | 19% | 16% | 20% | 27% |
| Pragmatists | 57%<br>*Intense*: 15%<br>*Relaxed*: 42% | 78% | 69% | 81% | 55% | 56% |
| Fundamentalists | 13% | 6% | 12% | 3% | 25% | 17% |
| Cynical Experts | 5% | | | | | |

*Figure 1*. Data analysis steps ©Elueze & Quan-Haase

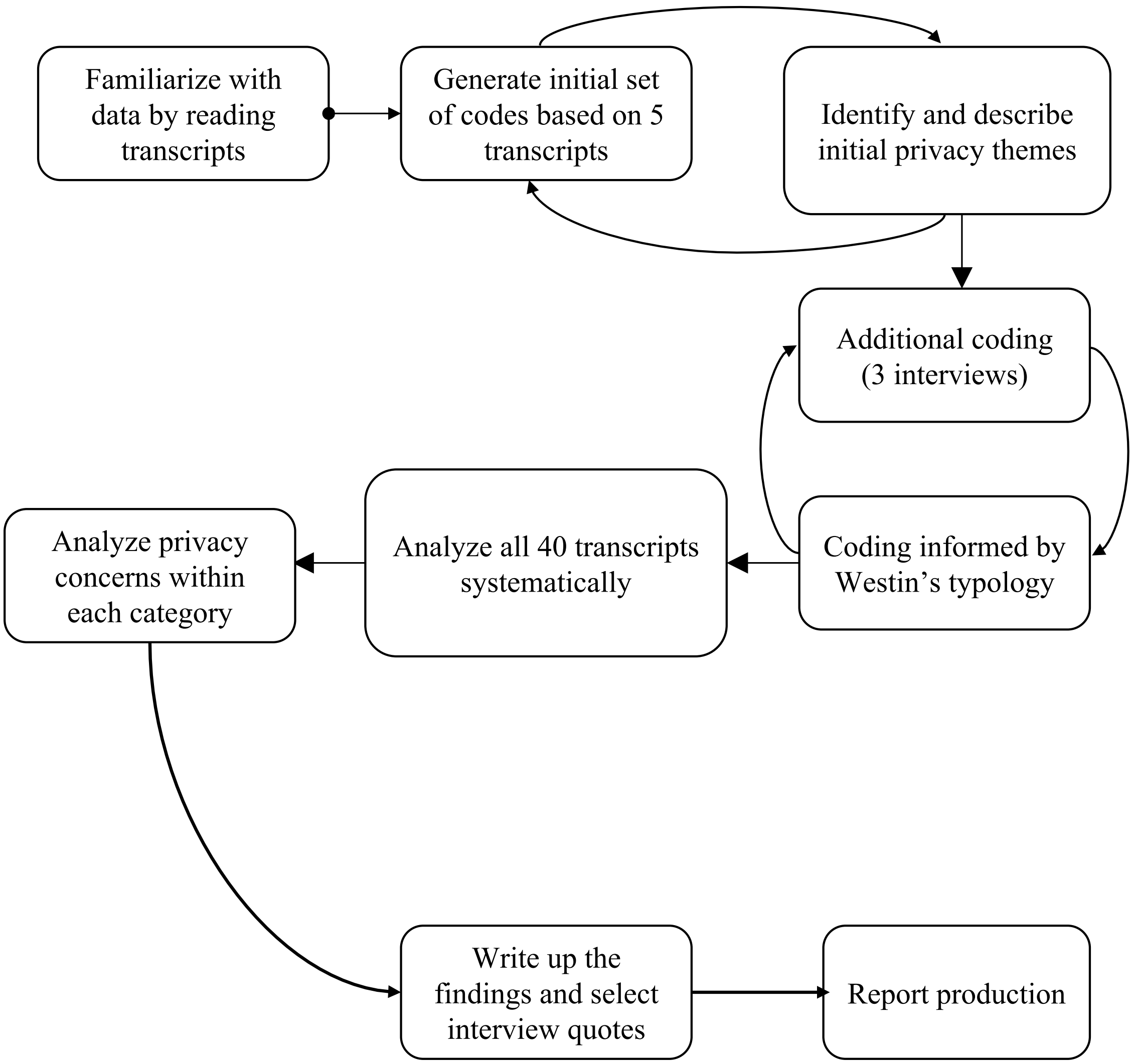

**Fundamentalists (13%)**

- Consider themselves very private
- Show paranoia regarding risks involved in disclosing personal information when using digital media
- Think people share too much information when using digital media (especially on social media)

**Marginally concerned (25%)**

- Do not consider themselves private
- Are not aware of the risks involved in information sharing online, or do not think their current online activities make them vulnerable to privacy risks
- Do not mind divulging information about themselves when using digital media

**Cynical Experts (5%)**

- Consider themselves private in some ways
- Well informed about privacy risks related to digital media
- Do not mind divulging information about themselves when using digital media
- Feel that an invasion of their privacy online and when using digital media is inevitable and beyond their control

**Intense Pragmatists (15%)**

- Consider themselves private in some ways
- Understand risks involved in disclosing information when using digital media
- Do not like having to disclose information, but demonstrate an understanding of the trade-off between protecting privacy and engaging with digital media, e.g., when making online purchases

**Relaxed Pragmatists (42%)**

- Consider themselves private in some ways
- Basic familiarity with the risks involved in sharing information when using digital media
- Do not like having to give out information when using digital media, but may do so depending on purpose, e.g., to use social media

*Figure 2*. Older adults' attitudes toward privacy by type (*N=40*) ©Elueze & Quan-Haase